# Increasing the Security of Advanced Metering Infrastructure Using Dynamic Defense Methods

Mohsen Saffar, Hamid Reza Naji
Dept. of Computer Engineering and Information Technology
Graduate University of Advanced Technology
Kerman, Iran
m.saffar@student.kgut.ac.ir, naji@kgut.ac.ir

***Abstract*— With the increase in Internet usage over the last half century, technology has also benefited from the growth and development of the Internet. It is no wonder that smart grids are emerging as one of the most popular technologies of today. As smart grids are connected to the internet, they are constantly vulnerable to cyberattacks. It is always possible for attackers to gain access to metering system infrastructure which is a major and critical part of smart grids. When this system is not working properly because of some security threats, irreparable damage is in the place for a country and its industries. This article suggests utilizing dynamic Applications that Participate in Their Own Defense defenses to secure such systems and infrastructures.**

## I. Introduction

In the 20th century, the Internet was one of mankind's greatest technological inventions. Internet development has been instrumental to global economic, social, health, control and energy development since it was born half a century ago and has significantly changed the lives of people. In spite of this, benefits and risks always coexist. The process of informatization is accompanied by cyberattacks like a nightmare that cannot be avoided. The development of human society has been affected by cyber security as a global issue [1].

The Smart Grid, known as the modern electric power grid has radically transformed the isolated and centralized energy supply system into a highly advanced and massively interconnected cyber-physical energy supply system that has benefitted from the rapid advances occurring in technology and the rapid adoption of IoT devices in the last two decades [2], [3].

In addition to being an essential resource for economic growth, energy is also a vital resource for all other important sectors of national infrastructure. It has become increasingly difficult to understand the operational vulnerabilities in a Smart grid. This is due to the different attack vectors that could threaten the stability and efficiency of the system. In addition, threats from nation-states and malicious actors are increasing. Among the most likely global threats over the next 10 years will be large-scale cyberattacks against critical infrastructure and networks, according to the World Economic Forum's 2020 Global Risks report. Kaspersky ICS-CERT reports the greatest number of vulnerabilities among all critical infrastructure sectors, according to reports from the US ICS-CERT and ICS-CERT. Various security risks can result from these cyber incidents, including the loss of sensitive data, malicious modifications, and deletions of critical power system states. Customers may receive inaccurate billing information, the energy market may be manipulated, and power outages of varying sizes may endanger lives by limiting power to other national critical infrastructure [4].

There are certain static characteristics of vulnerabilities that can be spotted by attackers in smart grid units. Embedded hardware environments, heterogeneous software environments, differentiated applications, and a variety of private communication protocols are all characteristics of power grid application environments, which makes it difficult to acquire a unified and standardized "acquired immunity" in order to defend against network attacks on power grids. In addition, static, similarity, and certainty provide attackers with an array of conveniences, including target recognition, detecting defensive behavior, improving attack technology tests, and assessing the effectiveness of the attack [5].

The limitations in the computing environment of smart grid units make it impossible for traditional protection measures to be applied. Despite the heterogeneity and diversity of smart terminal environments in power grids, such terminals have a large number of industrial applications that can be executed in heterogeneous enclosed environments that operate continuously and in a heterogeneous manner. The challenges that they face include limited computing power, high reliability requirements, complex mechanisms for destroying malicious programs, and complex protection technologies to contend with. In addition, they also face limited computing power. It is impossible for traditional terminal security protection products to be applied to smart terminals as a result of the complexity of the implementation and the high usage of resources [6].

Through advanced metering infrastructure (AMI), smart grid applications are able to learn and record electric power consumption on demand or on a schedule, which is the first step toward future smart grid implementation [7]. It can also be said that the AMI is one of the key components of the smart grid that is being implemented today, which allows for bi-directional communication between electric utility companies and their customers [8]. There are several different components that make up the AMI, including the information/communication network, smart meters, and meter data management software. AMI's network of communication is mainly composed of the home area network (HAN), the wide area network (WAN), and the utility system which are the three key components of the AMI's communication network. Currently, an AMI uses an access network structure that is based largely on a more traditional network approach. This approach is not always reliable and is prone to network failures, which is one of the most common weaknesses of this strategy [9], [2], [10].

Dynamic defense against attacks has the following effects on the attacker:

It has been observed that attackers have difficulty finding targets first, which causes them to fail; in the second place, attackers discover the wrong targets; in the third place, attackers discover targets but can't launch attacks; attackers are able to launch attacks, but they are unable to sustain them; and attackers in fifth place are capable of launching attacks, but they are easily detected by their targets [5], [11], [12].

As long as the above characteristics are present, any type of technology that meets the above criteria can be categorized as a dynamically enabled cyber security defense technology.

To achieve smart grid unit self-immune and universal immunity against grid application-level attacks specially, it is discussed in this paper how to use the application, Applications that Participate in Their Own Defense (APOD) in in AMI for smart grid units.

## II. MATERIAL AND METHOD

Up until now, static defense schemes have dominated efforts to protect the grid from malicious attacks. In contrast, in this paper, we will discuss a dynamic defense policy that is intended to counter network defenses. This policy addresses dynamic interactions between attackers and defenders in order to combat network attacks [13]. As a result of the dynamic nature of the internet, dynamic network defense refers to providing dynamic defense at the network layer. This includes several network elements, such as network topology, network configuration, network resources, network nodes and network services at the network layer. Dynamic network defense breaks staticty, certainty, and similarity of network elements by dynamization, virtualization, and randomization. By doing so, it increases the difficulty of network detection and intranet node penetration in order to prevent malicious attacks by hackers on a target network. Based on recent research, it has been found that the time of early investigation accounts for 95% of the overall process in the case of a cyberattack. The time of launching an attack accounts for only 5% of that process. The implementation of a dynamic network defense system that can interfere in the early detection and reconnaissance of attackers, even providing them with false information, will be an appropriate and effective way to defend the network [1].

In terms of technology, Applications That Participate in Their Own Defense (APOD) can be understood as a type of synchronization-based end-to-end information hopping protection technology [14]. It involves hopping end information between two sides of the communication chain to transmit data end-to-end. It is a technique by which both sides or one side of the communications chain changes the end information such as the ports, addresses, timeslots, encryption algorithms, or even protocol pseudo randomly according to the protocol, in order to prevent interference and attacks by enemies and achieve active network protection. This technique is referred to as end information hopping at the network layer, and it falls under node information hopping. Information is hopped between a server and a peer host on one side or between a server and a peer host on two sides depending on how the participants are categorized. An AV hybrid hopping defense policy was proposed in the APOD project, and a network protection system was developed. It is a method of anti-port scanning and anti-DoS (Denial of Service) attacks based on false port address hopping. DoS can lead to blackouts in smart grids that can be even more destructive than natural disasters [4], [15].

In China, Lin and Jia et al. From Nankai University proposed an active network protection system model based on end information hopping, which destroys attacker interference and achieves network protection by changing end information such as communication ports, addresses, time slots, and even protocols in end-to-end data transmission pseudo randomly. A prototype system for the end information hopping technique has been developed by Java mobile agents using an encryption algorithm for hopping and synchronization. This proves the technique can successfully resist DoS attacks [1].Moving target defense was suggested for smart grid by Ge et.al [3]. This method is the main dynamic defense approach[1]. Mimic dynamic defense was proposed for smart grids [5] they showed the benefits of dynamic defense for grids.

A partial successful attack can be mitigated by utilizing APOD, a dynamic network defense mechanism that complements and enhances other protective measures. When an application fails to function correctly, it is corrupt. Applications that are corrupt may deliver poor service or fail to deliver any service at all. Defending against the corruption of critical applications is therefore aimed at preventing or reducing the damage suffered by these applications. It is possible for an application to be corrupted for a number of reasons, including an accident, such as a hardware failure, or malicious intent. The reasons are either the lack of protection in its environment that allows it to be damaged, or the reasons why the application is misbehaving due to flaws in its own implementation. It is considered defense-enabled when an application has mechanisms in place preventing most attackers from corrupting it as fast as they can without these mechanisms, even if the application does not include that mechanism. Therefore, an attacker from a hacker community must not only defeat security mechanisms in the environment, but he must also spend additional time defeating defense mechanisms built into the application itself in order to be successful. It is important to understand that privilege plays a central role in both attacks and defenses. As a result of an attack, the attacker is able to gain access to some critical application that he can use to corrupt it. To be effective, defense enabling has to either prevent the attacker from being able to gain such privileges or keep him from being able to effectively use them once he has them [15].

Defense enabling, at its core, is not just about slowing down or preventing privileged attackers from acquiring privileges. Instead, it is about responding and adapting to privileged attackers' abuse of resources and responding to their attacks. The system is divided into several security domains, each with its own set of privileges, in order to slow the spread of privilege. By forcing the attacker to take more time accumulating privileges, the malicious application can be corrupted more slowly. As a result, the attacker does not have to spend more time collecting privileges by creating multiple security domains: for instance, if a particular domain were a single point of failure for the application, a malicious attacker could kill all application processes in that domain by gaining domain administrator privileges. A redundant distribution across the domains is clearly necessary. Replicating every essential part of the application across multiple domains is the simplest solution. When an attacker corrupts a single replica of the system, the problem of defense is transformed into a problem of fault tolerance. To ensure that the replicas will not be corrupted when some of them are corrupted, they must be coordinated so that they will not be corrupted collectively [15].

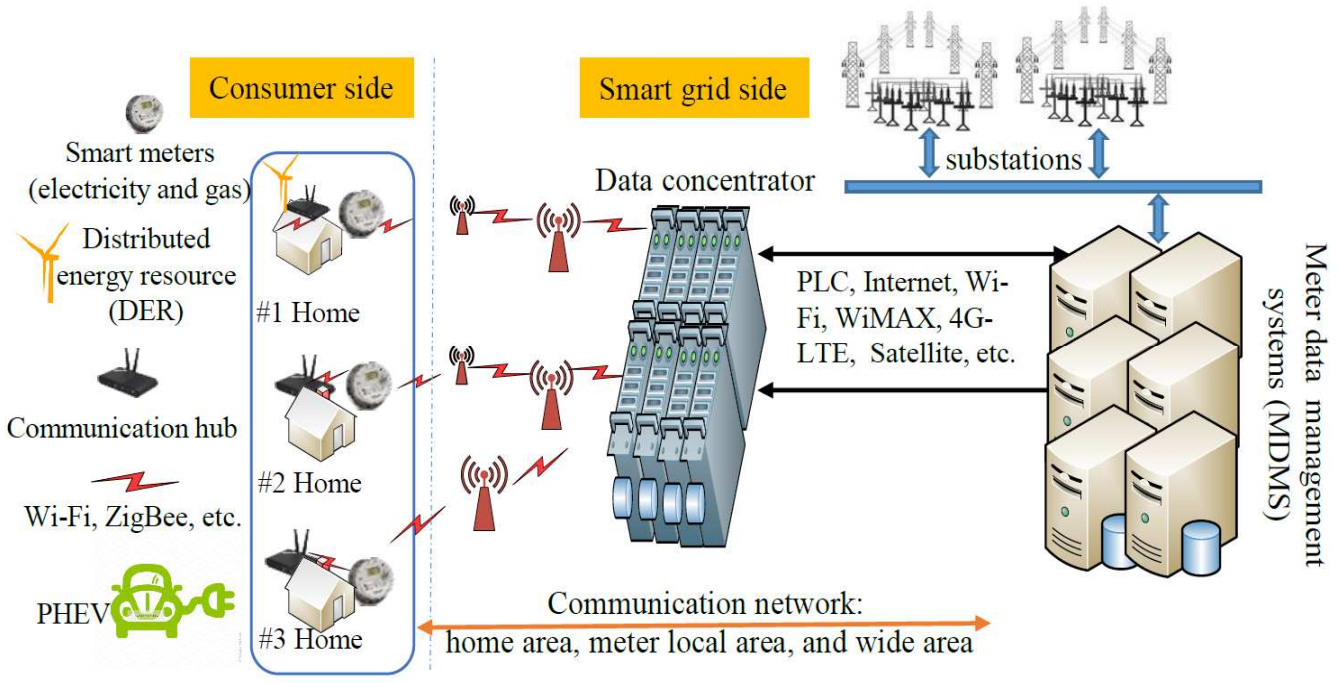

Figure1. AMI data transmission scheme based on [15].

A communication network is an enabling technology in order to realize the SG metering networks. As shown in Fig. 1, communication networks can be categorized into three types.

**Home Area Network (HAN)**: A HAN can be defined as a collection of appliances, entertainment systems, lighting systems, energy storage systems, and electric vehicles available to consumers [15].

Additionally, a home display and controller provide an interactive user interface for controlling energy consumption and maintaining devices. An energy smart meter acts as a home gateway for energy consumption measurement, sending collected information to the utility's control data center and executing control commands [15].

**Meter Local Area Network (MLAN):** Smart meters and data concentrators communicate through the meter local area network. Data concentrators are located in the energy distribution domain, and valves, switches, capacitor controllers, and monitors are located in the field area networks [15].

Powerline communication (PLC) is typically used by MLAN to send data between nodes using existing wireline connections [15].

**Wide Area Network (WAN):** Data concentrators and utility control centers communicate through a WAN, the highest-level network in the SG metering network. Through the WAN, massive amounts of data, commands, and signals are transferred and received for SMI applications. A number of technologies can be used to provide WAN communication, such as IEEE 802.16 (i.e., WiMAX), IEEE 802.20 (MobileFi), PLC, IEEE 802.11 (WiFi), and IEEE 802.15.4 (ZigBee) [15].

**Meter Data Management System (MDMS):** Communication technologies are typically used to collect information from smart meters in SG metering networks. Meter data management systems (MDMS) store these informations on utility servers [15].

Data collected from a power grid can be used for various purposes, such as network monitoring, operational management, and billing.

On the way to transforming traditional power grids into smart grids, advanced metering infrastructure (AMI) is a significant step. Smart meters, user panels, and communication infrastructure have been installed in many countries to implement AMI. A smart meter system consists of a communication network, data management systems, and a data collection system. Data on energy consumption is sent periodically or in real-time on demand by AMI's communication part to the monitoring centers through smart meters. In addition, it sends alerts about events to the control center. Users receive information on the power market and load control from control centers.

End-user devices use electronic hardware and software to collect and store data at desired intervals and time stamp it. A remote data center is in communication with these devices. By setting up time slots, the system administrator can distribute information to various parties at specified times. With AMI, smart devices or load controllers can accept commands thanks to bidirectional communication, which sets it apart from Automatic Meter Reading (AMR). Consumers and service providers can access consumption information through smart meters. Smart devices can be monitored by In-Home Displays (IHDs) through which consumers can view their energy consumption [16].

As a summary, electricity meters have the following characteristics:

- Pricing based on time.
- Data collection for consumer and utility consumption.
- Net metering.
- Notification of failures and outages.
- Commands (turning on/off) via remote control.
- A load limiting mechanism for using Demand Response.
- Monitoring of power quality, including voltage, current and phase, reactive and active power, power factor.
- Detection of energy theft.
- Interaction with other intelligent devices.
- Reducing emissions by using energy efficiently [17].

In order to achieve the smart grid's goals, ensuring the robustness of the AMI system's communication system is essential. AMI system vulnerabilities lead to various attacks, so it is crucial to recognize them. In addition, removing and detecting them correctly improves the reliability and efficiency of the system.

Sending and receiving information to/from the control centers is one of the most significant security goals of the AMI system.

In this case, it was discussed above that it is possible to use APOD as a security measure against DoS attacks, so, as I mentioned earlier, when the departmental communication was being established with each other, end information such as communication ports, addresses, slot times, and protocols used in end-to-end data transmission could be changed in a pseudo-random manner as dynamic defense of this smart grid system.

## III. DISCUSSION

The Internet is one of the greatest inventions until now. Benefits and risks of this technology have also grown as a result of its advancement. This has become a global problem and these risks have a significant effect on human societies. In modern power grid, known as the smart grid, a large number of physical cyber systems have been connected that allowing it to be more effective and stable in the face of many challenges that this technology presents, but also there are numerous vulnerabilities, which are causing it to become increasingly vulnerable. AMI is a network-based part of smart grid. There are various applications of smart grid technology, among which AMIs have taken an important first step towards the first phase of smart grid development. AMIs have enabled power companies to provide customers with two-way communication by reading and recording electricity consumption under a specific schedule or on demand, allowing power companies to meet customers' needs. According to reports, the energy sector has the highest number of vulnerabilities due to cyberattacks compared to other critical infrastructures. There are several problems associated with these systems, one of them being that static vulnerabilities are easily identified and exploited by attackers.

As an alternative to traditional attack defense methods, the dynamic defense method is recommended that will be used in the life cycle of a cyber information system. APOD, is an dynamic network defense mechanism that complements and enhances other protective measures. In this paper we suggested using the APOD method that involves hopping end information between two sides of the communication chain to transmit data end-to-end for AMI. It is a technique by which both sides or one side of the communications chain change the end information as a dynamic defense. In addition to the fact that this method reduces access for attackers due to dynamic access, it also creates a lengthy and time-consuming path for them in case of admission. Even with their access, it saves the mentioned address as a corrupted address, and for this reason, it prevents other addresses from connecting with it, and it will be another barrier against the access of attackers.

There are some factors that distinguish the APOD approach to the ability to survive from others. Dynamic adaptation is a central theme of this approach. Second, it is typically application and mission specific. These strategies are beyond traditional techniques for security. Third, the defense can be used in middleware as an intermediary between the application and the networks and operating systems. Recently developments in middleware have allowed us to adapt defenses quickly. So, it can easily be used for AMI in smart grids as a dynamic, fast defense [14].

## IV. CONCLUSION

In smart grids, essential parts of energy controllers by the internet are at attack risk and the traditional static defense method cannot protect them effectively. In this paper APOD as a dynamic network defense method is recommended for AMI in smart grid to increase the security and protection of these systems.